\documentclass[12pt,onecolumn]{IEEEtran}
\IEEEoverridecommandlockouts

\usepackage{amsmath}
\usepackage{booktabs}
\usepackage{makecell}
\usepackage{graphicx}
\usepackage{algorithm}
\usepackage{algpseudocode}

\title{Dataflow \& Tiling Strategies in Edge-AI FPGA Accelerators:\\
       A Comprehensive Literature Review}

\author{
    Zhaoqin Li\\
    Department of Electrical Engineering and Computer Science\\
    Samueli School of Engineering\\
    University of California, Irvine\\
    Irvine, CA, USA\\
    Email: zhaoqil3@uci.edu
}

\begin{document}

\maketitle
\thispagestyle{plain}
\pagestyle{plain}

\begin{abstract}
Edge-AI applications demand high-throughput, low-latency inference on FPGAs under tight resource and power constraints. This survey provides a comprehensive review of two key architectural decisions for FPGA-based neural network accelerators: (\emph{i})~the \emph{dataflow} (the order and manner in which data is moved and reused on chip), and (\emph{ii})~the \emph{tiling/blocking} strategy (how large tensors are partitioned to fit on-chip). We first present a broadened taxonomy of canonical dataflow styles—Weight-Stationary, Output-Stationary, Row-Stationary, and No-Local-Reuse—including formal definitions, pseudo-code/diagrams, and real FPGA examples. We then discuss analytical frameworks (\textit{MAESTRO}, \textit{Timeloop}) and compare them with a concise feature table, illustrating how they model reuse, performance, and hardware costs, and include a case study of a $3\times3$ convolution layer to demonstrate typical tool outputs. Next, we detail multi-level tiling and loop unrolling/pipelining strategies for FPGAs, clarifying how each memory tier (registers, LUTRAM, BRAM, HBM) can be exploited. Our four case studies—\textbf{FINN}, \textbf{FINN-R}, \textbf{FlightLLM}, and \textbf{SSR}—highlight distinct dataflows (from binary streaming to hybrid sparse transformations) and tiling patterns. We include a unified comparison matrix covering platform, precision, throughput, resource utilization, and energy efficiency, plus small block diagrams for each design. We conclude by examining design automation trade-offs among HLS, DSL, and hand-coded RTL, offering a “lessons learned” summary box, and charting future research directions in partial reconfiguration, hybrid dataflows, and domain-specific compiler flows for next-generation edge AI FPGA accelerators.
\end{abstract}


\section{Introduction}
Deploying \emph{deep neural networks} (DNNs) at the \emph{edge devices} demands high performance per watt under limited on-chip memory and power budgets. \emph{FPGAs} excel here by enabling customized parallelism, bitwidths, and memory hierarchies, achieving energy-efficient throughput compared to CPUs or GPUs~\cite{Umuroglu2017, Blott2018}. However, realizing these benefits requires \textbf{(1)~careful dataflow selection} to maximize on-chip reuse of weights and activations, and \textbf{(2)~optimal tiling/blocking} to match data shapes to limited BRAM/URAM/HBM.

In this paper, we:
\begin{enumerate}
    \item Present an expanded taxonomy of four \emph{canonical dataflow styles} for DNN accelerators on FPGAs (Section~\ref{sec:dataflow-taxonomy}).
    \item Compare \emph{analytical frameworks} (MAESTRO vs.~Timeloop) for data-centric modeling (Section~\ref{sec:frameworks}).
    \item Examine tiling, loop unrolling, pipelining, and multi-level buffering in detail (Section~\ref{sec:tiling}).
    \item Survey four case studies---FINN, FINN-R, FlightLLM, SSR---with a unified quantitative comparison (Section~\ref{sec:casestudies}).
    \item Assess design automation approaches (HLS vs.~DSL vs.~RTL) and highlight key productivity trade-offs (Section~\ref{sec:design-automation}).
    \item Conclude with \textbf{Future Directions} and a succinct \textbf{Conclusion} (Sections~\ref{sec:future}--\ref{sec:conclusion}).
\end{enumerate}

Readers seeking broad background on DNN architectures are referred to~\cite{Chen2016}. Our focus is on \emph{edge-tier FPGA} designs, from binarized CNNs to large language models.

\section{Dataflow Taxonomy}
\label{sec:dataflow-taxonomy}
Dataflow defines how the nested loops of a DNN layer (over input channels, output channels, spatial axes) are mapped onto hardware PEs and on-chip memory~\cite{Kwon2019}. We present four canonical styles below, each with a short pseudo-code or snippet, a small block diagram, and an example FPGA design.

\subsection{Weight-Stationary (WS)}
\label{sec:ws}
\textbf{Definition:} In Weight-Stationary dataflow, each \emph{PE} keeps its filter weights locally, and partial sums / activations stream through. Once weights are loaded, they remain `stationary` on-chip as inputs are cycled across them.

\begin{algorithm}
\caption{Weight-Stationary Dataflow}
\begin{algorithmic}[1]
\ForAll{tiles of input activations}
    \State Load tile into local buffer
    \ForAll{PE in array}
        \State \textit{partial\_sum} $\gets$ 0
        \ForAll{elements in tile}
            \State \textit{partial\_sum} $\gets$ \textit{partial\_sum} $+$ weight[PE] $\times$ activation[element]
        \EndFor
        \State output[PE] $\gets$ \textit{partial\_sum}
    \EndFor
\EndFor
\end{algorithmic}
\end{algorithm}

\textbf{Diagram:}

\begin{figure}[h]
    \centering
    \includegraphics[width=0.6\linewidth]{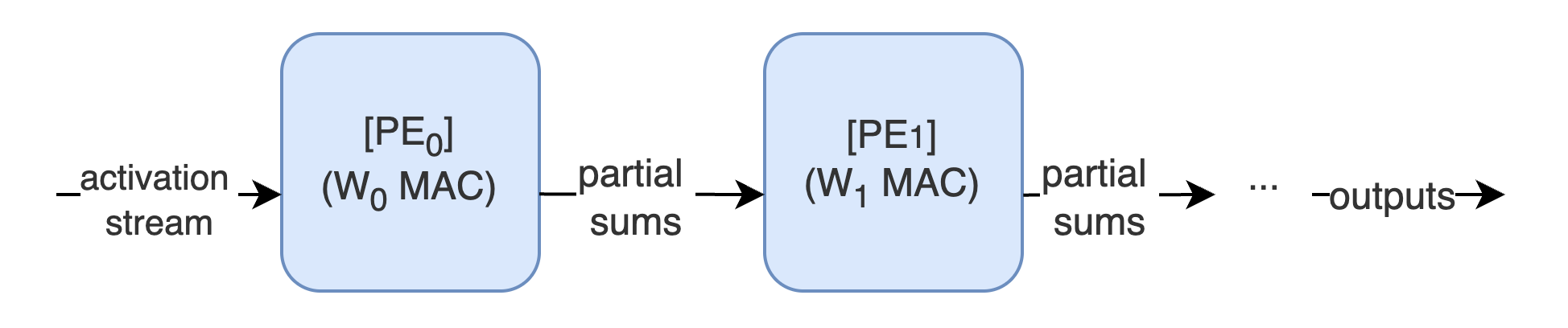}
    \caption{Weight-Stationary dataflow}
    \label{fig:enter-label}
\end{figure}

\textbf{Example FPGA Accelerator:}
Many systolic-array-like designs use WS. For instance, Google’s TPU v1 had a weight-stationary strategy in its 2D systolic array (though it is an ASIC). On FPGAs, some \emph{matrix-multiply} IP cores or HLS-based designs pin weights in BRAM to minimize re-fetching~\cite{Umuroglu2017}.

\subsection{Output-Stationary (OS)}
\label{sec:os}
\textbf{Definition:} In Output-Stationary, partial sums for a specific output element (or a small group of output channels) remain in local PE registers or buffers until all input channels are processed, after which the final result is written back to memory.

\begin{algorithm}[h]
\caption{Output Stationary GEMV-like Dataflow}
\begin{algorithmic}[1]
\For{$\text{out\_ch} \gets 0$ \textbf{to} Out}
    \State $\text{partial\_sum}[\text{out\_ch}] \gets 0$
\EndFor
\For{$\text{in\_ch} \gets 0$ \textbf{to} In}
    \For{$\text{out\_ch} \gets 0$ \textbf{to} Out}
        \State $\text{partial\_sum}[\text{out\_ch}] \gets \text{partial\_sum}[\text{out\_ch}] + \text{weight}[\text{out\_ch}, \text{in\_ch}] \times \text{input}[\text{in\_ch}]$
    \EndFor
\EndFor
\State \textbf{Write} $\text{partial\_sum}[]$ \textbf{to memory}
\end{algorithmic}
\end{algorithm}

\textbf{Diagram:}
\begin{figure}[h]
    \centering
    \includegraphics[width=0.6\linewidth]{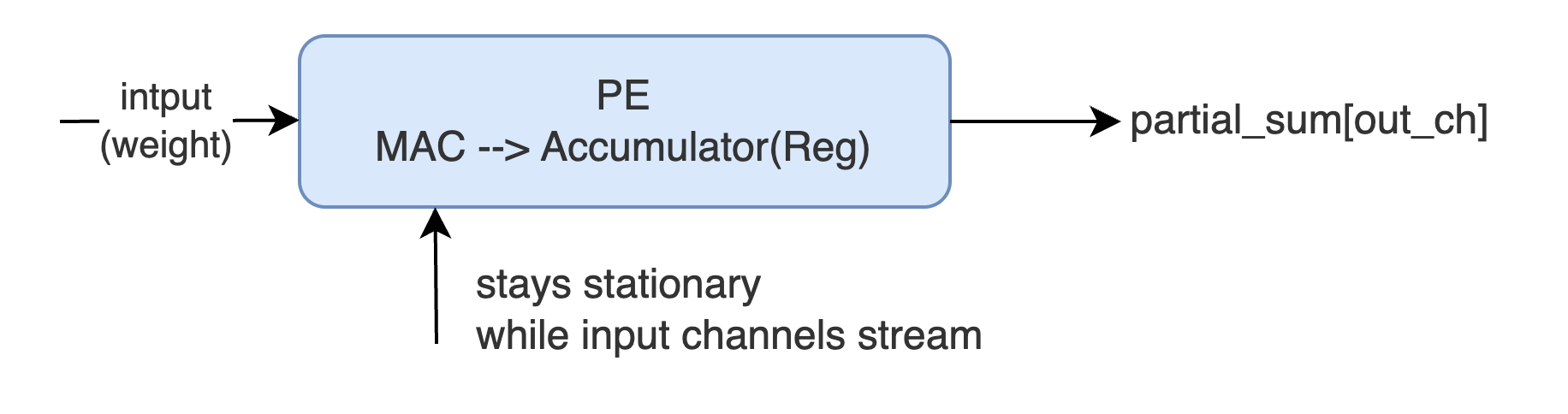}
    \caption{Output-Stationary dataflow}
    \label{fig:enter-label}
\end{figure}

\textbf{Example FPGA Accelerator:}
Some earlier FPGA CNN pipelines used OS by accumulating the result of one output pixel in local registers. ShiDianNao (an ASIC) also exemplifies OS, streaming input data across PEs that keep partial sums in place. Eyeriss (v1) effectively uses an OS-like scheme for many conv layers~\cite{Chen2016}.

\newpage
\subsection{Row-Stationary (RS)}
\label{sec:rs}
\textbf{Definition:} Proposed in Eyeriss~\cite{Chen2016}, Row-Stationary keeps a \emph{strip} (or row) of activations and partial sums on each PE, so that all relevant filters can be reused across them. It balances weight and activation reuse.

\begin{algorithm}
\caption{Row-wise Tiled Convolution with PE-Level Accumulation}
\begin{algorithmic}[1]
\ForAll{rows $r$ in output}
    \State Load row $r$ of input tile into \textit{row\_buffer}
    \ForAll{filter sets}
        \ForAll{processing elements (PEs) in row $r$}
            \State $\text{partial\_sum}[PE] \gets \text{partial\_sum}[PE] + \text{MAC}(\text{row\_buffer}[PE], \text{filter\_chunk})$
        \EndFor
        \State \dots (e.g., update filter chunk or control)
    \EndFor
    \State Store completed row $r$ to output memory
\EndFor
\end{algorithmic}
\end{algorithm}

\textbf{Diagram:}
\begin{figure}[h]
    \centering
    \includegraphics[width=0.4\linewidth]{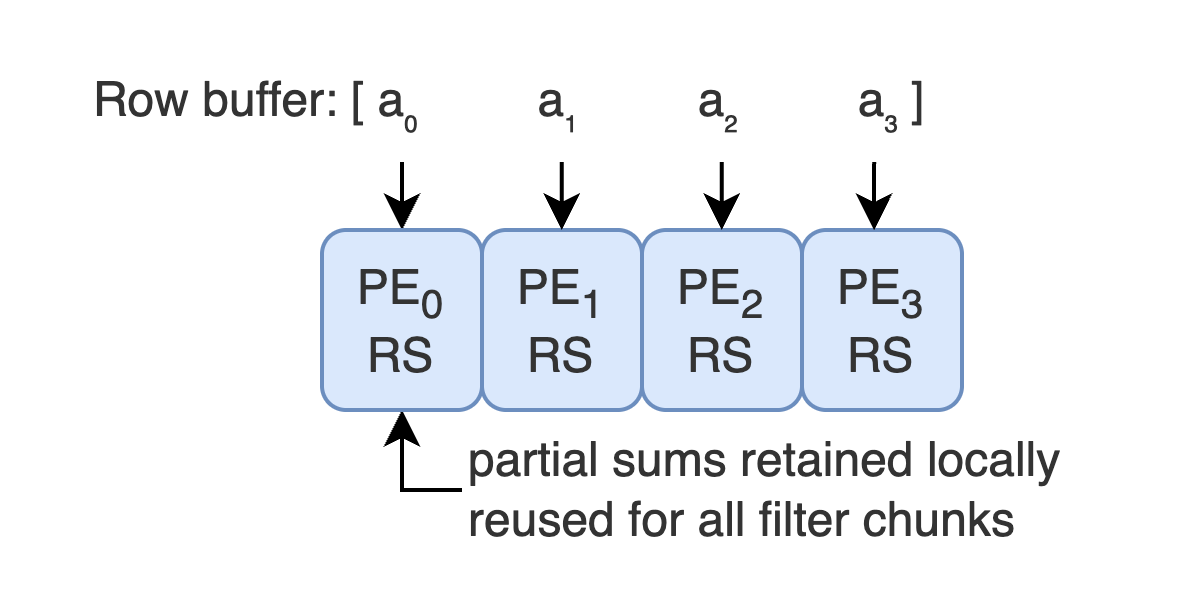}
    \caption{Row-Stationary dataflow}
    \label{fig:enter-label}
\end{figure}

\textbf{Example FPGA Accelerator:}
Eyeriss was first demonstrated as an ASIC. On FPGAs, \emph{row-stationary} is less common but still relevant where the cost of local reuse for both weights and partial sums is balanced. Some HLS-based designs adapt RS to minimize DRAM access by storing partial sums row-by-row in BRAM.

\subsection{No-Local-Reuse (NLR)}
\label{sec:nlr}
\textbf{Definition:} Also called “global-buffer reuse.” PEs have minimal local storage; data (weights, activations, partial sums) is read from a large shared buffer (or high-bandwidth memory) each time. This simplifies control but increases buffer bandwidth demand.

\begin{algorithm}
\caption{Element-wise Output Dataflow with Global Buffer Access}
\begin{algorithmic}[1]
\ForAll{elements $e$ in output}
    \State $\text{partial\_sum} \gets 0$
    \For{$\text{in\_ch} \gets 0$ \textbf{to} $I$}
        \State $\text{partial\_sum} \gets \text{partial\_sum} + \text{global\_buffer.read}(\text{in\_ch}) \times \text{global\_buffer.read}(\text{weight\_idx})$
    \EndFor
    \State $\text{global\_buffer.write}(\text{output\_idx}, \text{partial\_sum})$
\EndFor
\end{algorithmic}
\end{algorithm}

\textbf{Diagram:}
\begin{figure}[h]
    \centering
    \includegraphics[width=0.4\linewidth]{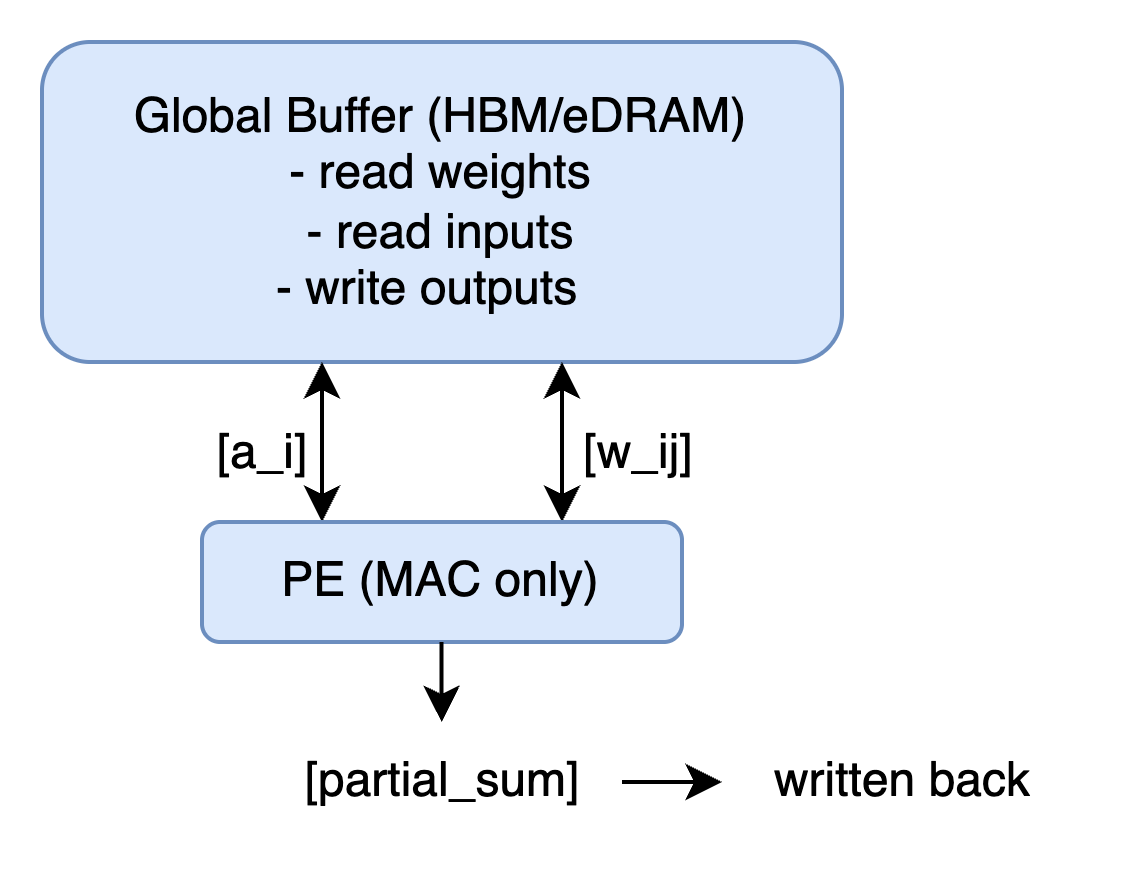}
    \caption{No-Local-Reuse Dataflow}
    \label{fig:enter-label}
\end{figure}

\textbf{Example FPGA Accelerator:}
DianNao (ASIC) used a large eDRAM “global buffer,” with minimal caching in the PEs. On FPGAs, some streaming matrix-multiply cores adopt a “flat read” from HBM + immediate compute + write-back.

\section{Analytical Frameworks for Dataflow \& Reuse}
\label{sec:frameworks}

Two widely used analytical frameworks for modeling DNN dataflow on custom accelerators are \emph{MAESTRO}~\cite{Kwon2020} and \emph{Timeloop}~\cite{Parashar2019timeloop}. Both tools systematically explore loop order, tiling factors, and data reuse patterns to estimate cycle count, energy consumption, and on‐chip buffer usage.

\paragraph{Comparison.}
\begin{table}[h]
  \centering
  \small
  \caption{Comparison of MAESTRO and Timeloop analytical frameworks}
  \label{tab:frameworks}
  \begin{tabular}{p{4.0cm}p{6.0cm}p{6.0cm}}
    \toprule
    \textbf{Attribute}                & \textbf{MAESTRO}                                                                           & \textbf{Timeloop}                                                                                                  \\
    \midrule
    Reuse directives                  & Data-centric DSL with explicit spatial/temporal reuse directives                           & Loop-nest enumeration with auto-search for optimal tiling~\cite{Parashar2019timeloop}                               \\
    Output metrics                    & Bandwidth, cycles, energy, buffer accesses per level~\cite{Kwon2020}                        & Cycles, energy, per-level buffer usage~\cite{Parashar2019timeloop}                                               \\
    Mapping specification             & User must provide tiling/unrolling factors for each buffer level~\cite{Kwon2020}            & User specifies architecture and constraints; mapper auto-explores tilings~\cite{Parashar2019timeloop}             \\
    Open-source availability          & Yes (C++ analysis core + Python API)~\cite{Kwon2020}                                       & Yes (C++ mapper + YAML spec; integrates with Accelergy)~\cite{Wu2019}                                            \\
    Simulation runtime                & $\approx0.5\,\mathrm{s}$ per network on a modern CPU (analytical only)~\cite{Kwon2020}      & Fast per-mapping evaluation; full-space searches scale with strategy~\cite{Parashar2019timeloop}                \\
    Extensibility                     & Fixed two-level hierarchy (PE scratchpad + global buffer); uniform PE clusters only~\cite{Kwon2020} & Arbitrary memory hierarchies, custom NoC and component models via plugins~\cite{Parashar2019timeloop}           \\
    Quantization \& sparsity support  & Fixed precision (e.g.\ 16-bit); no built-in sparse modeling~\cite{Kwon2020}                & Mixed-precision workloads; “Sparseloop” extension for sparse dataflow modeling~\cite{Wu2021}                     \\
    Integration into design flows     & Used in DSE loops (genetic/RL mappers) for HW/SW co-design~\cite{Wu2019}                    & Standalone pre-RTL evaluation; interfaces with DSL compilers or HLS flows via Accelergy~\cite{Wu2019}          \\
    Modeling granularity              & Analytical energy/area per access (PE, SRAM, NoC); configurable cost parameters~\cite{Wu2019} & Technology-aware energy/area modeling (per-bit SRAM, MAC, wire) through Accelergy~\cite{Wu2019}               \\
    \bottomrule
  \end{tabular}
\end{table}

\paragraph{Case Study: 3$\times$3 Convolution Layer Analysis.}
To illustrate typical results, we analyze a $3\times3$ convolution from ResNet-50~\cite{He2016} (64 input $\rightarrow$ 64 output channels on $56\times56$ feature maps, batch 1) on a hypothetical 256-PE accelerator with a two-level memory (256\,kB global buffer, 1\,kB PE-local buffer) using a weight-stationary dataflow.

MAESTRO reports:
\begin{itemize}
  \item \textbf{Weight reuse:} $\approx3{,}136\times$ (each $3\times3$ filter remains local and is applied across all spatial positions)~\cite{Kwon2020}.
  \item \textbf{Activation reuse:} up to $576\times$ via spatial broadcast and temporal accumulation~\cite{Kwon2020}.
  \item \textbf{Off-chip traffic:} $\sim0.87\,\mathrm{MB}$ (weights + activations) versus $440\,\mathrm{MB}$ if no on-chip reuse~\cite{Kwon2020}.
  \item \textbf{Latency:} $\approx5\times10^{5}$ cycles (ideal double buffering, full PE utilization)~\cite{Kwon2020}.
  \item \textbf{Energy breakdown:} dominated by on-chip MAC and SRAM accesses; DRAM accounts for $\ll5\%$ of total energy~\cite{Kwon2020}.
\end{itemize}

Timeloop (with Accelergy) yields comparable metrics:
\begin{itemize}
  \item \textbf{Cycle count:} $\approx$4.5\,$\times$10$^{5}$ cycles~\cite{Parashar2019timeloop}.
  \item \textbf{Energy profile:} $\sim$2\% DRAM energy, remainder in PEs and on-chip buffers~\cite{Parashar2019timeloop}.
  \item \textbf{Mapping exploration:} identifies weight-stationary as energy-optimal, with energy variance up to 11$\times$ across alternative tilings~\cite{Wu2021}.
\end{itemize}

This case study demonstrates how both frameworks expose data reuse, bandwidth requirements, and performance/energy trade-offs \emph{prior} to any HLS or RTL implementation, guiding accelerator designers toward near-optimal dataflow strategies and avoiding costly rework.  
\section{Tiling \& Blocking Strategies in FPGA Design}
\label{sec:tiling}
Efficient tiling and loop transformations are central to implementing dataflows on FPGAs with limited on-chip memory. This section dissects: (1)~Two-Level Tiling, (2)~Loop Unrolling \& Pipelining, (3)~Buffer Hierarchies \& Double Buffering, and (4)~Multi-Level Tiling for deeper memory tiers.

\subsection{Two-Level Tiling}
\label{sec:two-level-tiling}
\textbf{Definition and Rationale:} Large DNN layers typically exceed on-chip capacity, so we \emph{block} data into \(\textit{outer tiles}\) (resident in DRAM/HBM) and \(\textit{inner tiles}\) (resident in BRAM/LUTRAM). Each inner tile is processed fully before fetching the next, reducing external memory accesses~\cite{Blott2018}.

\begin{itemize}
    \item \textbf{Outer tile size} is chosen so that a large chunk can be loaded into on-chip buffers once and reused many times.
    \item \textbf{Inner tile size} is tuned to the PE-array shape (e.g., \(16\times16\)) so that the data can be consumed in a single pipeline.
\end{itemize}

\emph{Guidelines for Choosing Tile Sizes}:
\begin{itemize}
    \item \emph{Fit in BRAM}: The tile must not exceed available BRAM or URAM capacity.
    \item \emph{Utilize DSPs fully}: Align tile dimension with the number of parallel multipliers to minimize underflow.
    \item \emph{Amortize loading cost}: A tile that is too small may incur overhead from repeated DRAM transfers.
\end{itemize}

\subsection{Loop Unrolling \& Pipelining}
\label{sec:loopunroll}

\textbf{Unrolling:} By duplicating hardware for inner-loop iterations, throughput increases linearly with the unroll factor, constrained by LUT/DSP availability.

\begin{algorithm}[H]
\caption{Loop Unrolling with Factor 16}
\begin{algorithmic}[1]
\State \textbf{\#pragma HLS UNROLL factor=16}
\For{$i \gets 0$ \textbf{to} \textit{tile\_size} \textbf{exclusive}}
    \State Perform parallel operations on 16 elements
\EndFor
\end{algorithmic}
\end{algorithm}

\textbf{Pipelining:} Achieving \(\textit{Initiation Interval (II)} = 1\) means each loop iteration can start every cycle, overlapping compute.

\emph{Expected II Improvements}: 
\[
\text{II}_{\text{naive}} > 1 
\quad \xrightarrow[\text{resolving deps}]{} \quad 
\text{II}_{\text{optimized}} = 1
\]
Careful memory partitioning and accumulation logic (e.g., tree reductions) help reach II = 1.

\subsection{Buffer Hierarchies \& Double Buffering}
\label{sec:bufferhierarchy}

Modern FPGAs offer multiple on-chip tiers:
\begin{enumerate}
    \item \textbf{Registers} / \textbf{LUTRAM}: Minimal capacity, but single-cycle access.
    \item \textbf{BRAM}: Larger capacity (tens to hundreds of kbits) with 1--2 read ports.
    \item \textbf{UltraRAM}: Embedded blocks with deeper capacity (on some devices).
    \item \textbf{HBM} (High-Bandwidth Memory): On-package DRAM with \(\sim\!300+\) GB/s.
\end{enumerate}

\textbf{Double Buffering} is crucial: for tile \(\textit{n}\), while it is being processed by the PE array, tile \(\textit{n+1}\) is fetched from external memory into a spare buffer. This overlap hides memory latency.

\begin{algorithm}[H]
\caption{Double Buffering Flow}
\begin{algorithmic}[1]
\While{true}
    \State Load tile\_{$n+1$} into \textit{buffer\_B}
    \State Compute tile\_{$n$} from \textit{buffer\_A}
    \State Swap(\textit{buffer\_A}, \textit{buffer\_B})
\EndWhile
\end{algorithmic}
\end{algorithm}

\subsection{Multi-Level Tiling}
\label{sec:multilevel-tiling}
Beyond two levels, one can exploit a deeper hierarchy: e.g., \emph{outer tiles in HBM}, \emph{mid-level tiles in BRAM}, and \emph{inner sub-tiles in registers or LUTRAM}. This is increasingly relevant on devices like Xilinx Versal or Alveo U280, which combine HBM, URAM, and distributed RAM. Typically:
\begin{itemize}
    \item The \emph{outermost tile} is sized to the HBM capacity and bandwidth.
    \item \emph{Intermediate tiles} are staged in URAM/BRAM for partial reuse.
    \item The \emph{innermost sub-tile} is streamed through the PE pipeline with minimal buffering in LUTRAM/registers.
\end{itemize}
By carefully partitioning data at each level, accelerators can minimize off-chip traffic and keep the PE array continuously busy.

\section{Case Studies of Edge-AI FPGA Accelerators}
\label{sec:casestudies}

To illustrate how different dataflows and tiling strategies combine in real designs, we review four exemplary accelerators: 
\textbf{(1)~FINN}, 
\textbf{(2)~FINN-R}, 
\textbf{(3)~FlightLLM}, 
\textbf{(4)~SSR}. 
Table~\ref{tab:unified-comparison} then unifies their key metrics (platform, precision, throughput, resource usage, energy efficiency). Small block diagrams highlight each design's data movement.

\subsection{FINN: BNN Streaming}
Proposed by Umuroglu \emph{et al.}~\cite{Umuroglu2017}, \textbf{FINN} accelerates \emph{binarized} convolutional networks. Each layer is mapped as a streaming pipeline with \textbf{weight-stationary} dataflow (the 1-bit weights stay in LUT arrays), while 1-bit activations flow through in a sliding-window manner. Intermediate layer outputs are kept on-chip in shift registers or small BRAM. 

\[
\text{Input} 
\;\xrightarrow{\text{Sliding Window}}\;
\text{XNOR+POPCOUNT}
\;\xrightarrow{\text{Bitwise Accum}}\;
\text{Output}
\]

Due to full pipeline parallelization and minimal bitwidth, FINN achieves high frame rates (10k+ fps on CIFAR-10) on modest FPGAs.

\subsection{FINN-R: Automated QNN Framework}
\textbf{FINN-R}~\cite{Blott2018} generalizes FINN to 2--8\,bit quantized networks and provides an automated toolflow. Designers specify throughput/resource goals, and FINN-R’s cost model chooses unroll factors, tiling parameters, and layer pipelines. Dataflow remains \emph{streaming} with partial weight-stationary blocks. Example usage:
\subsection{FlightLLM: Sparse DSP for Large Language Models}
\textbf{FlightLLM}~\cite{Zeng2024} targets Transformers on Alveo U280 with \emph{HBM} plus configurable “sparse DSP” chains. Weights are partially pruned; dense blocks are loaded from HBM, but zeros are skipped. The \emph{dataflow} is partially \emph{output-stationary} for each token’s vector, combined with weight-streaming from HBM. 

\subsection{SSR: Spatial-Sequential Hybrid on Versal}
\textbf{SSR}~\cite{Zhuang2024} exploits \emph{multi-engine parallelism} on Xilinx Versal ACAP. It partitions Transformer layers across multiple accelerators, each using a mixture of on-chip AIEs for matrix multiply and PL for layernorm or activation. Dataflow is \emph{row-stationary} within each AIE cluster, with cross-layer partial sums passed in on-chip buffers. 

\subsection{Quantitative Comparison}
\label{sec:quant-comparison}

\begin{table*}[h]
\centering
\caption{Unified Comparison of Four FPGA Accelerator Case Studies}
\label{tab:unified-comparison}
\begin{tabular}{lcccccc}
\toprule
\textbf{Accelerator} & \textbf{Platform} & \textbf{Precision} & \textbf{Key Feature} 
& \textbf{Throughput} & \textbf{Resource Util.} & \textbf{Energy Eff.} \\
\midrule

\textbf{FINN}~\cite{Umuroglu2017} 
& \makecell[l]{Zynq-7020\\ (28nm)} 
& \makecell[c]{1-bit\\ weights/activ.} 
& \makecell[l]{Streaming,\\ layer-pipelined} 
& \makecell[c]{$\sim$5~TOPS\\ bin ops} 
& \makecell[l]{$\sim$50k LUT (90\%)\\ 0 DSP} 
& Not reported \\

\textbf{FINN-R}~\cite{Blott2018} 
& \makecell[l]{Xilinx VU9P\\ (16nm)} 
& \makecell[c]{2--8\,bit QNN} 
& \makecell[l]{Automated\\ unroll/tiling}
& \makecell[c]{50~TOPS\\ (1\,bit)} 
& \makecell[l]{$\sim$900k LUT (75\%)\\ 0 DSP} 
& $>$100~GOPS/W \\

\textbf{FlightLLM}~\cite{Zeng2024} 
& \makecell[l]{Alveo U280\\ + HBM (16nm)}
& \makecell[c]{8\,bit\\ + sparsity} 
& \makecell[l]{Sparse DSP\\ chaining} 
& \makecell[c]{$\sim$55 tokens/s\\ on LLaMA-7B} 
& \makecell[l]{574k LUT (44\%)\\ 6345/9024 DSP (70\%)} 
& 300--400~GOPS/W \\

\textbf{SSR}~\cite{Zhuang2024} 
& \makecell[l]{Versal VCK190\\ (7nm ACAP)}
& 8\,bit 
& \makecell[l]{Spatial-seq\\ multi-engine} 
& \makecell[c]{Up to 26.7~TOPS\\ int8} 
& \makecell[l]{394/400 AIE used;\\ PL partial} 
& 246--453~GOPS/W \\
\bottomrule
\end{tabular}
\end{table*}

\newpage
\section{Design Automation Approaches}
\label{sec:design-automation}
Accelerating DNNs on FPGAs can be done via \emph{HLS}, \emph{domain-specific languages (DSL)}, or hand-coded \emph{RTL}. Table~\ref{tab:automation-compare} contrasts these approaches, focusing on code size, compile times, typical resource overhead, and productivity gains.

\begin{table}[h]
\centering
\caption{HLS vs.\ DSL vs.\ RTL: Side-by-Side Comparison}
\label{tab:automation-compare}
\begin{tabular}{lccc}
\toprule
& \textbf{HLS} & \textbf{DSL} & \textbf{RTL (HDL)} \\
\midrule
\textbf{LoC (approx.)}      & Medium       & Very low     & High       \\
\textbf{Compile Time}       & Hours        & Hours--days  & Days--weeks\\
\textbf{Productivity}       & Medium       & High         & Low        \\
\textbf{Overhead (LUT/DSP)} & +10--30\%    & +20--40\%    & Baseline   \\
\bottomrule
\end{tabular}
\end{table}

\noindent
\begin{center}
\fbox{\begin{minipage}{.97\linewidth}
\textbf{Lessons Learned (Callout Box)}:
\begin{itemize}
    \item \textbf{Automation Gains:} HLS/DSL can reduce design time \(\approx 5\times\) vs.\ RTL, at a cost of \(\sim\)30\% more LUT usage.
    \item \textbf{Parallelism Leverage:} Automated flows still let you specify unroll factors; not purely “push-button.”
    \item \textbf{Verification Ease:} DSL-level simulation is faster; verifying each cycle in RTL is time-consuming.
    \item \textbf{Future Trend:} Hybrid approaches combining DSL for coarse architecture with selective hand-tuned kernels.
\end{itemize}
\end{minipage}}
\end{center}

\section{Future Directions}
\label{sec:future}
\begin{itemize}
    \item \textbf{Hybrid Dataflow Mappings:} Future accelerators may dynamically switch between weight-stationary and output-stationary (or row-stationary) depending on layer geometry and sparsity.
    \item \textbf{Partial Reconfiguration:} FPGAs can reconfigure part of the fabric to tailor each layer’s dataflow or tiling. Overheads must be balanced with potential gains.
    \item \textbf{Domain-Specific Compilers:} Building on MAESTRO/Timeloop, next-gen DSLs could unify code generation, loop transformations, and memory mapping for diverse networks (CNN, LSTM, Transformer) with minimal user intervention.
\end{itemize}

\section{Conclusion}
\label{sec:conclusion}
We presented a comprehensive overview of \emph{dataflow} styles and \emph{tiling} strategies for FPGA-based DNN accelerators. By dissecting canonical patterns—Weight-Stationary, Output-Stationary, Row-Stationary, No-Local-Reuse—and comparing frameworks like MAESTRO vs.~Timeloop, we illustrated how to systematically exploit on-chip reuse. Our deep dive into tiling (two-level and multi-level) and loop transformations underscores the complexity in balancing memory capacity, bandwidth, and DSP usage. Four case studies (FINN, FINN-R, FlightLLM, SSR) reveal success in tackling different model sizes and workload constraints, highlighting design automation’s growing importance. Looking ahead, partial reconfiguration, hybrid dataflows, and more sophisticated compilers promise to further shrink the gap between flexible FPGA solutions and the aggressive performance demands of edge AI.

\section*{Acknowledgment}
The author gratefully acknowledges insightful discussions with researchers from the \textbf{FINN}, \textbf{FlightLLM}, and \textbf{SSR} teams. This work was supported in part by the \textit{UC Irvine Edge AI Initiative}.

Special thanks to \textbf{Professor Pramod Khargonekar}, Distinguished Professor of Electrical Engineering and Computer Science at UC Irvine, for his guidance, encouragement, and continued support of edge AI research and education.

\bibliographystyle{IEEEtran}

\end{document}